\documentclass[conference]{IEEEtran}
\IEEEoverridecommandlockouts
\usepackage{cite}
\usepackage{amsmath,amssymb,amsfonts}
\usepackage{algorithmic}
\usepackage{graphicx}
\usepackage{placeins}
\usepackage{booktabs}
\usepackage{multirow}
\usepackage{textcomp}
\usepackage{xcolor}
\usepackage{colortbl}
\usepackage[colorlinks=true,allcolors=blue]{hyperref}
\def\BibTeX{{\rm B\kern-.05em{\sc i\kern-.025em b}\kern-.08em
    T\kern-.1667em\lower.7ex\hbox{E}\kern-.125emX}}

\newcommand{\runin}[1]{\par\noindent\textbf{#1.}\ \ignorespaces}

\begin{document}
\bstctlcite{IEEEexample:BSTcontrol}

\title{\textbf{FRAPPE}: \textbf{F}ull Input, \textbf{R}esidual Output \textbf{A}utoencoding with \textbf{P}rojection \textbf{P}ursuit \textbf{E}ncoder}

\author{\IEEEauthorblockN{Dan Jacobellis}
\IEEEauthorblockA{\textit{Dept. of Electrical and Computer Engineering} \\
\textit{The University of Texas at Austin}\\
Austin, TX 78712, USA \\
danjacobellis@utexas.edu \\
ORCID: 0000-0001-8541-1906}
\and
\IEEEauthorblockN{Neeraja J. Yadwadkar}
\IEEEauthorblockA{\textit{Dept. of Electrical and Computer Engineering} \\
\textit{The University of Texas at Austin}\\
Austin, TX 78712, USA \\
neeraja@austin.utexas.edu \\
ORCID: 0009-0007-7556-3069}
}

\maketitle

\begin{abstract}
Media compression standards have reached a plateau in terms of the rate-distortion-complexity trade-off, limiting the ability to offload expensive AI perception to the cloud in applications like robotics, wearables, and remote sensing. DNN-based codecs improve compression efficiency, but at a cost: they cannot easily adapt to large changes in available bitrate, and real-time encoding requires expensive, power-hungry GPUs that prohibit use on low-cost or resource-constrained platforms. To address these limitations, we propose a novel autoencoding framework (FRAPPE) that uses the \textbf{F}ull input to predict the \textbf{R}esidual output via a \textbf{P}rojection-\textbf{P}ursuit \textbf{E}ncoder. FRAPPE's encoding objective naturally sorts latent channels by importance, allowing zero-overhead variable-rate coding. Unlike RNN-based learned codecs, whose encoder consumes the previous reconstruction's residual, or RVQ-style codecs, whose codebooks must be applied sequentially, FRAPPE's analysis path is an embarrassingly parallel DAG of independent input projections. Using FRAPPE, we build a variable-rate RGB image codec (FRAPPE-Image), and evaluate its rate-distortion-complexity trade-off against standard image codecs. At high compression ratios ($\sim$0.1\,bpp) FRAPPE-Image provides higher perceptual quality than AVIF with 47$\times$ faster encoding, making it capable of real-time 1080p, 30fps CPU-only encoding. Our code and pre-trained models are available: \url{https://github.com/UT-SysML/FRAPPE}.
\end{abstract}


\section{Introduction}

Current media compression standards like VVC and AV1 have reached a plateau in terms of the rate-distortion-complexity trade-off.
Since the standardization of digital media codecs like JPEG and MP3 more than three decades ago, codec design innovations have led to dramatic improvements in signal quality for the same bitrate.
However, these increasingly complex designs are burdened by equally dramatic increases in encoding cost and power consumption~\cite{bossen2021vvc}.
For this reason, simpler codecs like JPEG and MP3 remain ubiquitous for power-constrained sensors~\cite{hojjat2025mcucoder}.
For many applications, particularly those involving robotics or wearables, this has severely limited the ability to offload computation to the cloud. Existing codecs all fail in at least one of three ways.
(1) They require prohibitively high encoding resources (FLOPS, memory bandwidth, etc.).
(2) They provide inadequate compression ratios to transmit data over the cellular, satellite, or BLE communication channels available in the field.
(3) They introduce too much distortion or latency to benefit from cloud-based processing.
Recent advances in deep neural network (DNN)–based autoencoders \cite{jacobellis2025learned, jacobellis2026liveaction} have shown potential to break free of this plateau, but make significant compromises in at least one of three dimensions:
(1) on-the-fly rate adaptation comparable to standards like JPEG or AVIF;
(2) encoding cost competitive with standardized codecs at matched compression efficiency;
(3) real-time encoding on commodity hardware without GPU or NPU accelerators for standard-resolution audio, image, or video streams.

To address these issues and improve the utility of cloud-assisted robotics and wearable applications, we propose a new type of residual autoencoder (FRAPPE).
FRAPPE uses the \textbf{F}ull input to predict the \textbf{R}esidual output via a \textbf{P}rojection-\textbf{P}ursuit \textbf{E}ncoder.
By using a projection pursuit encoding scheme, FRAPPE sorts the latent channels by importance, allowing zero-overhead variable-rate and progressive coding using a single set of encoder weights.
Unlike RNN-based learned codecs~\cite{toderici2015variable, toderici2017full}, whose encoder consumes the previous reconstruction's residual at each iteration, making them prohibitively expensive, and RVQ-style codecs~\cite{zeghidour2021soundstream, defossez2023high, kumar2023high}, whose codebooks must be applied sequentially, FRAPPE formulates the residual autoencoding objective using the \textit{full input} to predict the \textit{residual output}. This decouples the per-channel projections so the analysis path becomes a DAG: all latent channels are encoded in parallel and the encoder collapses to $S$ strided convolutions at inference, without any recurrence or quantizer chain. Our contributions are threefold.
\begin{itemize}
    \item We propose FRAPPE, an autoencoding framework designed to provide (1) variable rate and progressive compression, (2) competitive rate distortion performance at high compression ratios, and (3) low encoding costs to enable use with resource constrained sensors 
    \item Using this framework, we instantiate and train a practical image compression system.
    \item We evaluate FRAPPE-Image against other conventional and learned image codecs and demonstrate extreme gains in terms of the rate-distortion-complexity trade-off.
\end{itemize}

% contributions are blah blah, highlight results

\runin{Background and related work} FRAPPE builds upon previous works related to asymmetric neural codec design, residual autoencoding, and projection pursuit algorithms.

\runin{Asymmetric neural codec design} The asymmetric design philosophy of WaLLoC~\cite{jacobellis2025learned} and LiVeAction~\cite{jacobellis2026liveaction}---a heavy nonlinear synthesis transform paired with a deliberately lightweight analysis transform---is well suited to resource-constrained encoding. FRAPPE inherits this stance, along with the log-variance rate proxy used by LiVeAction. MCUCoder~\cite{hojjat2025mcucoder} achieves encoding efficiency gains in an asymmetric architecture using post-training quantization.

\runin{Residual autoencoding} The closest neural-codec analogues are the Toderici et al.\ recurrent compressors~\cite{toderici2015variable, toderici2017full}, which encode an image as a chain of additive reconstructions $\hat{x}_t = \hat{x}_{t-1} + D_t(E_t(r_{t-1}))$ in which each stage's encoder consumes the previous reconstruction's residual, requiring the decoder to be evaluated inside the encoding loop. Neural audio codecs built on residual vector quantization~\cite{zeghidour2021soundstream, defossez2023high, kumar2023high} avoid this by pushing the residual recursion into the quantizer chain instead, but the chain itself remains sequential at encode time. More broadly, fitting a sum of terms one at a time on the residual of the preceding fit is the forward stagewise additive modeling framework~\cite{hastie2009elements_ch10_3}, of which projection-pursuit regression is the supervised special case. Classical signal-processing precursors include matching pursuit~\cite{mallat1993matching} and orthogonal matching pursuit~\cite{pati1993orthogonal}---greedy dictionary expansions whose atom-selection rule is itself recognized as a special case of projection pursuit~\cite{pati1993orthogonal}---alongside the cascade-correlation constructive network~\cite{fahlman1990recurrent} and greedy layer-wise autoencoder pretraining~\cite{bengio2006greedy}, which add components one at a time but operate on a latent rather than an output residual. Closest in spirit to FRAPPE's deflation pattern are alternating least squares for nonlinear PCA~\cite{young1978principal, michailidis1998gifi}, deflation-based canonical correlation analysis~\cite{knapp1978canonical, hardle2015applied_16}, and one-unit deflation-mode FastICA~\cite{hyvarinen1997fast, hyvarinen1999fast}, the last of which explicitly identifies each extracted direction with a projection-pursuit index.

\runin{Projection pursuit} Projection pursuit~\cite{friedman1974projection, hastie2009elements_ch11_2} is a family of methods for finding informative linear projections $\hat{k}^\top X$ of multivariate data by varying the projection direction $\hat{k}$ so as to maximize a continuous index of ``usefulness.'' In the original algorithm formulation~\cite{friedman1974projection}, unconstrained hill-climbing is applied to a smoothed index measuring the product of global spread and local density in the projected dimension, producing multiple distinct projections by restarting from different seeds and constraining subsequent searches to subspaces orthogonal to already-found directions.  Projection pursuit regression~(PPR)~\cite{friedman1981projection} extends the method to supervised learning by fitting the additive model
  \begin{equation}
    f(X) = \sum_{m=1}^{M} g_m(\omega_m^\top X),
    \label{eq:ppr}
  \end{equation}
  where each $\omega_m$ is a learned unit projection direction and each $g_m$ is a nonlinear function.
  PPR is fit forward-stagewise: at stage $m$, a new pair $(\omega_m, g_m)$ is added to minimize the residual error left by the previous $m-1$ components, and prior directions are typically
  frozen~\cite{hastie2009elements_ch11_2}.
  The number of components $M$ is determined by the stagewise procedure itself: fitting terminates when the next term no longer appreciably improves the fit.

\section{Proposed Method}

To enable real-time, cloud-assisted machine perception on the resource-constrained sensors used in robotics and wearables, FRAPPE is designed around three goals:
(1) zero-overhead variable-rate and progressive coding with a \emph{single} set of encoder weights;
(2) rate--distortion performance competitive with standardized codecs (JPEG, AVIF);
(3) high-throughput encoding on low-power sensors without GPUs or accelerators.

\runin{Codec workflow} Let $x\in\mathbb{R}^{C\times T_1\times\cdots\times T_D}$ denote a signal with $C$ channels and $D\in\{1,2,3\}$ spatio-temporal dimensions, normalized to $[-1,1]$. FRAPPE composes an analysis transform $\mathcal{G}_{\!A}$, an entropy-coded quantizer $\mathcal{Q}$, and a synthesis transform $\mathcal{G}_{\!S}$:
\begin{equation}
\hat{x} \;=\; \mathcal{G}_{\!S} \,\circ\, \mathrm{Adapt}_{p_d} \,\circ\, \mathcal{Q} \,\circ\, \Phi \,\circ\, \mathcal{G}_{\!A}(x).
\label{eq:pipeline}
\end{equation}
The analysis transform $\mathcal{G}_{\!A}$ splits into $S$ scale groups, where group $s$ carries $n_s$ latent channels at patch size $p_s$; each channel is a single learned linear projection of a non-overlapping patch of $x$. The companding nonlinearity $\Phi$ confines every channel to a signed 8-bit range; the quantizer $\mathcal{Q}$ rounds to integers and per-scale latents are entropy-coded independently. Before reconstruction, $\mathrm{Adapt}_{p_d}$ rebins each scale's grid to a common decoder resolution $p_d$ and the resulting tensors are concatenated for $\mathcal{G}_{\!S}$. The trained instance evaluated in Section~\ref{sec:results} (henceforth FRAPPE-Image) operates on RGB images ($C{=}3$, $D{=}2$) and uses $S{=}5$ scale groups with $(n_s,p_s)$ $=$ $(3,32)$, $(6,16)$, $(3,8)$, $(6,4)$, $(3,2)$ for $N{=}21$ latent channels total, and $p_d{=}8$.

\runin{(a) Residual autoencoding with a progressively relaxing entropy bottleneck} FRAPPE introduces channels one at a time in coarse-to-fine order. Let $\mathcal{F}_{m-1}$ denote the merged codec over the first $m{-}1$ channels (with $\mathcal{F}_0\!\equiv\!0$). The $m$-th channel's encoder--decoder pair is fit to the \emph{output-space} residual $r_m = x - \mathcal{F}_{m-1}(x)$ by minimizing
\begin{equation}
\begin{split}
\mathcal{L}_m \;=\; & \log_{10}\!\bigl\lVert r_m - \hat r_m \bigr\rVert_2^2 \\
& {} + \lambda_m\,\bigl(\mathbb{E}\,r_m^2\bigr)^{\rho}\,\log_2 \mathrm{Std}\!\bigl(\Phi(\omega_m^{\!\top}\mathrm{Patch}_{p_m}(x))\bigr),
\end{split}
\label{eq:loss}
\end{equation}
where $\omega_m$ is the new channel's projection direction, $\hat r_m$ is the single-channel autoencoder's prediction, and the second term is LiVeAction's log-variance rate proxy~\cite{jacobellis2026liveaction} re-weighted by the (detached) residual power $\mathbb{E}\,r_m^2$ raised to $\rho{=}0.3$; without this re-weighting the rate term grows to dominate the distortion term as residual energy decays, collapsing later channels to near-zero output.

\begin{figure}[t]
\centering
\includegraphics[width=\columnwidth]{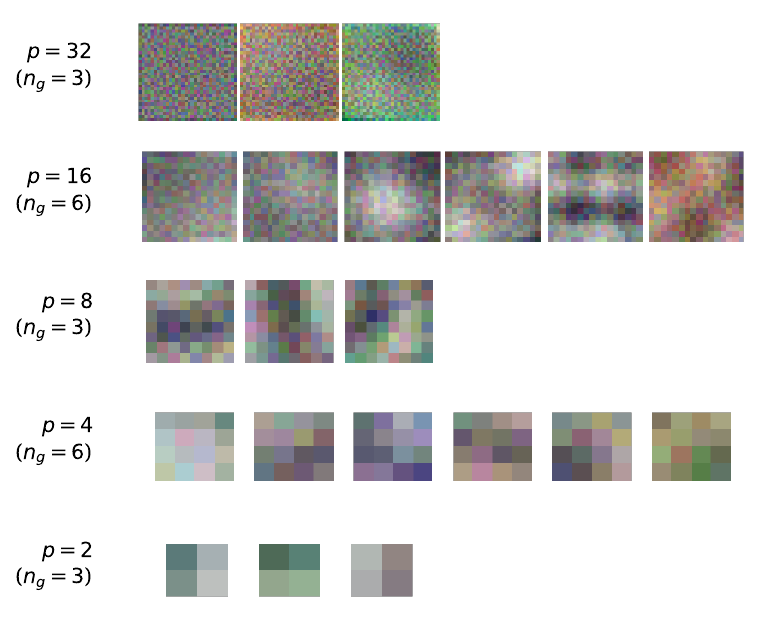}
\caption{Consolidated encoder weights of FRAPPE-Image, one row per scale group. Each tile is a learned filter $\omega_m\in\mathbb{R}^{3\times p_s\times p_s}$ rendered as RGB, normalized to $\pm 4\sigma$ within its scale row. The five rows show $(n_s,p_s)=(3,32),(6,16),(3,8),(6,4),(3,2)$ for $N{=}21$ channels. When trained on sRGB inputs, FRAPPE-Image learns, without supervision, a representation similar to chroma subsampling in a luma, chrominance-orange, chrominance-green (YCoCg) color space.}
\label{fig:filters}
\end{figure}

\begin{figure}[t]
\centering
\includegraphics[width=\columnwidth]{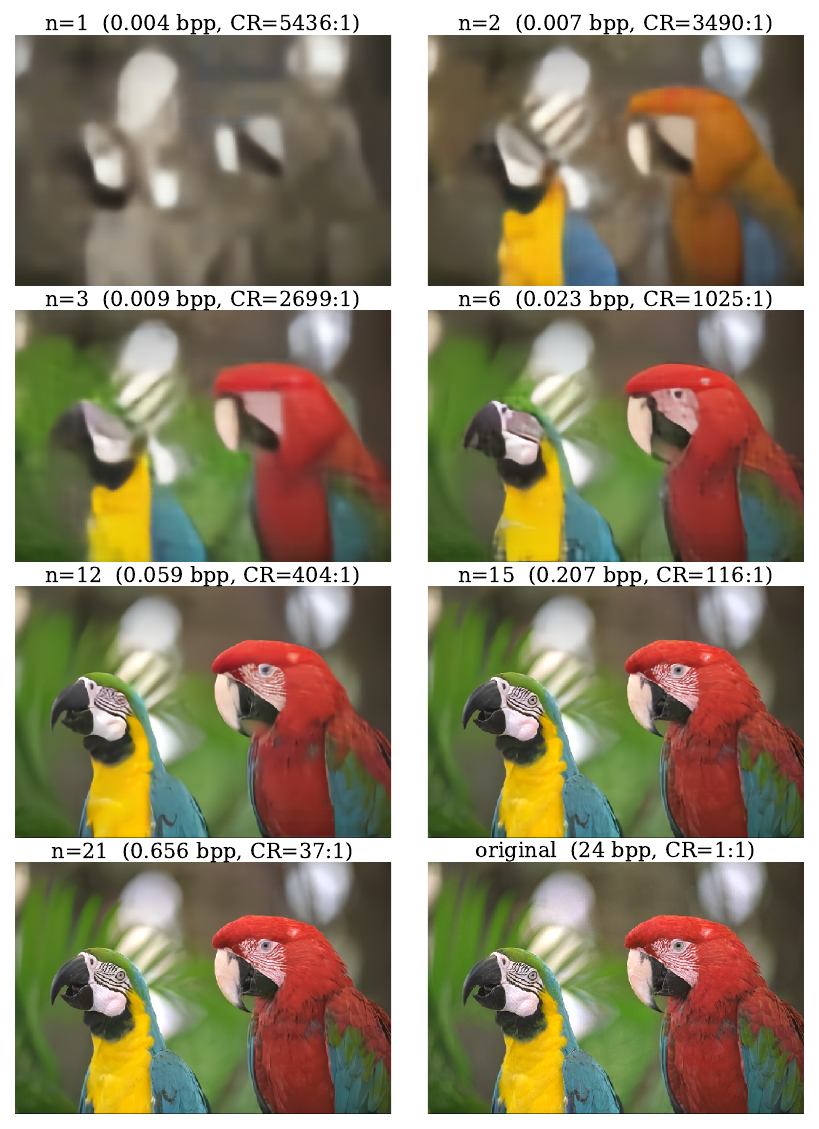}
\caption{Progressive reconstructions of \texttt{kodim22} as the transmitted channel count $n$ is varied. All $n$ panels share the same encoder weights; only the truncated channel count and matching merged-decoder snapshot differ. The bottom-right panel is the uncoded reference. Bits-per-pixel measurements are JPEG-LS-coded.}
\label{fig:progressive}
\end{figure}

The patch size and the Lagrangian $\lambda_m$ relax monotonically across the sequence. Channel~0 has the most aggressive bottleneck: at patch size $p_0$ a single 8-bit latent samples a $C p_0^D$-dimensional input patch, paired with the largest $\lambda_m$ (for FRAPPE-Image, $3{\times}32^2{=}3072$, a $3072{:}1$ per-channel dimensionality reduction). By the final channel the bottleneck has relaxed to $C p^D$ at the finest patch size and a smaller $\lambda_m$ (for FRAPPE-Image, channel~20 at $p_{20}{=}2$ gives $12{:}1$). Each new channel only needs to capture variance not already explained by its predecessors, so the schedule yields latents that are naturally sorted by importance with no explicit decorrelation loss. Fig.~\ref{fig:filters} shows the resulting filters: coarse channels carry low-frequency luma/chroma DC, finer scales sort into oriented edges and color textures. This intrinsic ordering directly delivers goal~(1): retaining only the first $n$ channels and selecting the matching merged-decoder snapshot recovers a rate point on the operating curve, with no auxiliary scale-selection module, fine-tuning, or encoder rerun. Fig.~\ref{fig:progressive} traces this sweep on a single Kodak image, and Fig.~\ref{fig:rd_throughput} (Section~\ref{sec:results}) reports the full curves over the Kodak set.

\runin{(b) Asymmetric design via full-input projection-pursuit encoding} DNN-based autoencoders earn their rate--distortion advantage by leveraging large datasets and substantial decode-time compute. FRAPPE targets an asymmetric deployment topology: capture-side encoding on resource-constrained sensors, cloud-side decoding on workstation hardware. This inverts the encode-once/decode-many model of broadcast media (VVC, AV1, HEVC), where decoder cost is the binding constraint; here it is paid once per upload at the cloud, which can transcode to formats suitable for downstream applications. We therefore adopt the asymmetric philosophy of WaLLoC~\cite{jacobellis2025learned} and LiVeAction~\cite{jacobellis2026liveaction}---a powerful nonlinear synthesis transform paired with a deliberately lightweight analysis transform---and combine it with the residual scheme of (a) by a specific design choice: each channel's encoder operates on the \emph{full input} $x$ rather than on the latent-space residual of previous channels. The training target stays the output-space residual $r_m$, but the encoder input does not.

This realizes the projection-pursuit regression model of Section~I (cf.~Eq.~\eqref{eq:ppr}). Channel~$m$ takes
\begin{equation}
z_m \;=\; \Phi\!\bigl(\omega_m^{\!\top}\,\mathrm{Patch}_{p_m}(x)\bigr),
\label{eq:single_channel}
\end{equation}
with $\omega_m\in\mathbb{R}^{C\,p_m^D}$ a learned projection direction and the per-channel ridge function $g_m$ realized jointly by the merged synthesis transform across all channels. Because all $n_s$ channels in scale group $s$ share the same patch size and the same input, their projections consolidate into a single strided convolution at inference,
\begin{equation}
z^{(s)} \;=\; \Phi\!\bigl(W^{(s)} \ast_{p_s} x + b^{(s)}\bigr),
\quad W^{(s)}\!\in\!\mathbb{R}^{n_s\times C\times p_s^{(D)}},
\label{eq:ga_consolidated}
\end{equation}
where $p_s^{(D)}$ denotes the $D$-fold product $p_s\!\times\!\cdots\!\times\!p_s$ and $\ast_{p_s}$ is $D$-dimensional strided convolution with stride $p_s$. The consolidation is exact---the channels were trained one at a time but never share a kernel or interact across channels in the analysis path---so the FRAPPE-Image encoder is just $S{=}5$ \texttt{Conv2d} layers followed by per-channel companding and quantization.

The synthesis transform absorbs nearly all the model's parameters and FLOPs. Its architecture is fixed across channel counts (only the first pointwise projection's input widens with $n$), but its weights are snapshotted: one retrained $\mathcal{G}_{\!S}$ per supported channel count. Because all encoders are frozen during this retraining, encoder weights are bit-identical across snapshots, and a single set of encoder weights serves every $n$. The body is a kernel-3 projection to a fixed width, a stack of ConvNeXt-style~\cite{liu2022convnet} residual blocks (depthwise kernel-3, $\mathrm{LayerNorm}$, pointwise expand by $4{\times}$, $\mathrm{GELU}$, pointwise contract, with LayerScale), a pointwise projection to $C p_d^D$ channels, a stride-$p_d$ transposed $D$-dimensional convolution, and $\mathrm{Hardtanh}$; FRAPPE-Image instantiates the stack at width 768 with twelve blocks. Each scale group's quantized latents are first remapped to the decoder resolution $p_d$,
\begin{equation}
\mathrm{Adapt}_{p_d}\bigl(z^{(s)}\bigr) \;=\; \begin{cases} \mathrm{S2D}_{p_d/p_s}\!\bigl(z^{(s)}\bigr), & p_s<p_d,\\[2pt] z^{(s)}, & p_s=p_d,\\[2pt] \mathrm{NN}_{p_s/p_d}\!\bigl(z^{(s)}\bigr), & p_s>p_d, \end{cases}
\label{eq:adapt}
\end{equation}
where $\mathrm{S2D}_f$ folds $f^D$-sample blocks into the channel dimension (one encoder channel becomes $f^D$ decoder channels) and $\mathrm{NN}_f$ is nearest-neighbor upsampling. The adapted tensors are concatenated and fed to $\mathcal{G}_{\!S}$. With $C_d = \sum_{p_s\le p_d}\! n_s(p_d/p_s)^D + \sum_{p_s>p_d}\! n_s$ adapted decoder-input channels (for FRAPPE-Image, $C_d=3+6+3+24+48=84$), \eqref{eq:pipeline} expands to
\begin{equation}
\hat{x} \;=\; \mathcal{G}_{\!S}\!\Bigl(\,\bigoplus_{s=1}^{S}\mathrm{Adapt}_{p_d}\!\bigl(\mathcal{Q}\,\Phi(W^{(s)}\!\ast_{p_s}\!x + b^{(s)})\bigr)\Bigr),
\label{eq:gs}
\end{equation}
with $\bigoplus$ denoting channel-wise concatenation.

\runin{(c) Cheap, parallelizable analysis transform} Because the analysis path consists only of a strided convolution and a pointwise nonlinearity, its per-sample cost is closed-form. The strided convolution from $C$ input channels to $N$ latent channels touches each input sample exactly once and contributes $CN$ multiply--accumulates per sample \emph{regardless of patch size}---a patch of size $p_s^D$ requires $Cp_s^D$ MACs but covers $p_s^D$ samples, so the per-sample cost is $C$ MACs per channel. The softsign compander $\Phi_c(u)=r u/(\sigma_c+|u|)$ with $r{=}127$ guarantees $|\Phi_c(u)|<r$ and so fits the companded activations into a signed 8-bit range. The denominator scale $\sigma_c$ is learned per latent channel, and a learned per-channel multiplier is applied to the output (one scalar each per channel); together they cost $4$ operations per latent element (absolute value, addition, division, post-softsign multiply; the fixed scalar $r$ fuses into the divide). Per sample, scale group $s$ contributes only $4 n_s/p_s^D$ companding ops. The full analysis path therefore costs $CN + \sum_s 4n_s/p_s^D$ ops per sample, dominated by the linear projection and independent of decoder depth or number of scale groups; for FRAPPE-Image this evaluates to ${\approx}68$ ops/pixel, with even the finest scale ($n_s{=}3$, $p_s{=}2$) adding just $3$ ops/pixel.

Equally important, the per-scale strided convolutions in \eqref{eq:ga_consolidated} share the input $x$ but are otherwise independent, so the analysis path forms an unconstrained DAG whose nodes can be pipelined or evaluated in parallel---there is no recurrent encoder dependency~\cite{toderici2015variable, toderici2017full} and no sequential residual-quantizer chain~\cite{zeghidour2021soundstream, defossez2023high, kumar2023high} to serialize the encode pass. After companding and rounding, scale group $s$ produces $\mathcal{Q}(z^{(s)})\!\in\!\mathbb{Z}^{n_s\times T_1/p_s\times\cdots\times T_D/p_s}$, which is serialized per scale and concatenated into the full bitstream; per-scale coding is the natural choice given that scales have different spatial resolutions. Pre-quantization activations approximately follow a generalized Gaussian distribution close to a Laplacian, as is typical of subband coefficients of natural signals~\cite{sharifi1995estimation}, so any 8-bit lossless codec whose prediction residuals are modeled with a Laplacian-like distribution is nearly entropy-optimal. The implementation isolates entropy coding behind a four-function contract so any modality-appropriate lossless codec can be substituted (e.g.\ FLAC for 1D signals); FRAPPE-Image reshapes each scale to a single 2D grayscale plane $(n_s\!\cdot\!T_1/p_s,\,T_2/p_s)$ and applies length-prefixed JPEG-LS~\cite{weinberger2000loco}, whose Golomb--Rice prediction residuals are two-sided geometric---the discrete analog of a Laplacian.

\runin{Training and implementation details} We train FRAPPE-Image on the LSDIR dataset~\cite{li2023lsdir} with batch size 1 using the Adan optimizer~\cite{xie2024adan}; Kodak is held out for validation. Each channel passes through two stages. The single-channel residual stage fits $(\omega_m,g_m)$ at peak learning rate $1.5{\times}10^{-5}$ on a steep cosine ramp; the small peak reflects that the encoder is being adapted to a residual that $\mathcal{G}_{\!S}$ already partially explains. After fitting, the new encoder weights are merged into their scale group, all $m$ encoders are frozen, $\mathcal{Q}$ is switched from training-time additive noise to hard rounding, and only $\mathcal{G}_{\!S}$ is retrained on the union of latents (with $\lambda{=}0$) at peak learning rate $5{\times}10^{-4}$ on a milder ramp. Within either stage the encoder parameter group runs at one-tenth the decoder learning rate, keeping the lightweight projections stable while the heavier synthesis transform absorbs most of the optimization signal. Per-channel epoch counts ramp coarse-to-fine (single-channel $2{\rightarrow}7$, merged-decoder $4{\rightarrow}7$), reflecting that later channels carry smaller residual energy and finer detail. The full per-channel $\lambda_m$ schedule and training scripts are available in the accompanying code repository\footnote{\url{https://github.com/UT-SysML/FRAPPE}}.

\section{Experimental Data and Results}
\label{sec:results}

\begin{figure*}[!tp]
\centering
\includegraphics[width=\textwidth]{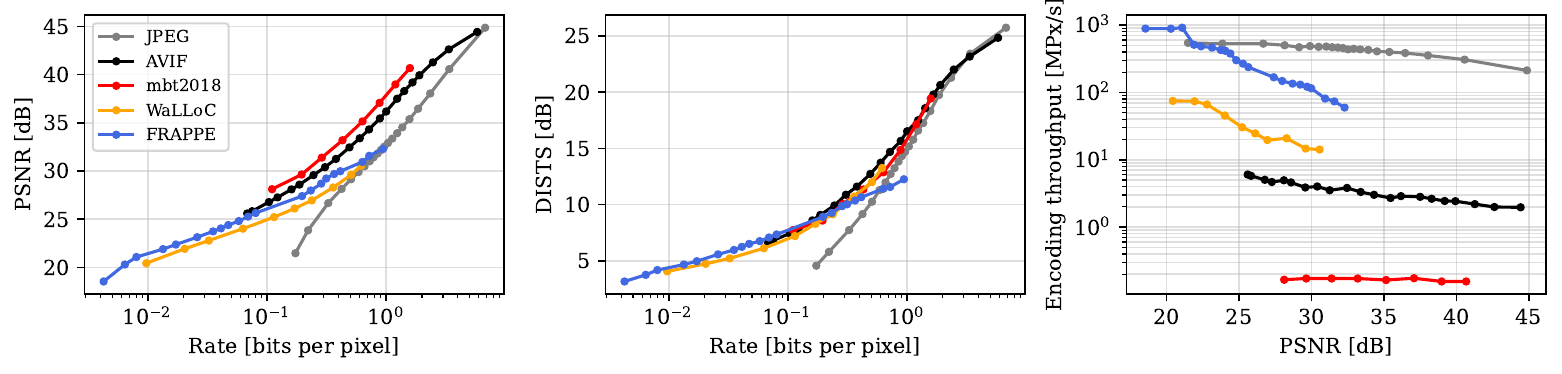}
\caption{Rate-distortion and encoding-throughput comparison on Kodak for JPEG and AVIF (both via Pillow), mbt2018 (via CompressAI), WaLLoC, and FRAPPE-Image. Left and middle: PSNR and DISTS vs rate (bits per pixel, log scale). Right: encoding throughput (MPx/s, log scale) vs PSNR.}
\label{fig:rd_throughput}
\end{figure*}

We evaluate the rate-distortion-complexity performance of FRAPPE-Image on the Kodak dataset. We compare against conventional transform codecs (JPEG, AVIF) as well as symmetric and asymmetric neural codecs (mbt2018~\cite{minnen2018joint} and WaLLoC~\cite{jacobellis2025learned}, respectively) on a shared CPU testbed (AMD EPYC 9354). Rate is measured using bits per pixel (bpp), where 24 bpp corresponds to 8-bit RGB inputs. Distortion is measured using conventional and perceptual metrics (PSNR, SSIM~\cite{wang2004image}, and DISTS~\cite{ding2020image}) at the original image resolution of 768$\times$512 or 512$\times$768. Following~\cite{jacobellis2025learned}, DISTS is reported in decibels as $\mathrm{DISTS_{dB}}\!=\!-10\log_{10}(\mathrm{DISTS})$ so that higher values indicate better perceptual quality. Consistent with FRAPPE's asymmetric deployment topology (Section~II), we report encoder-side throughput, measured as the median over five timed runs (one warmup) on CPU and timed end-to-end through the analysis transform, companding/quantization, and JPEG-LS entropy coding; no GPUs or hardware accelerators are used at inference for any codec. AVIF results use Pillow over libavif at default speed and effort with no tile or thread tuning---the configuration most production deployments rely on. Fig.~\ref{fig:rd_throughput} compares the rate--distortion--complexity trade-off of FRAPPE-Image against JPEG, AVIF, mbt2018, and WaLLoC; additional measurements against JPEG XL, LiVeAction~\cite{jacobellis2026liveaction}, and MCUCoder~\cite{hojjat2025mcucoder} are reported in the appendix, with FRAPPE-Image holding a $+3.1$ to $+4.3$\,dB BD-PSNR lead over MCUCoder at matched rate.

\runin{Exceptional performance at high compression ratios} At bitrates near 0.1\,bpp (compression ratio of 240:1) FRAPPE-Image provides better perceptual quality (DISTS) than AVIF and 47 times faster encoding. The advantage extends across the low-rate band: FRAPPE attains the best mean BD-DISTS in every regime below 0.215\,bpp against every baseline in Table~\ref{tab:bd_metric}.

\runin{Real-time CPU-only encoding} FRAPPE-Image is capable of real-time (1080p, 30fps) CPU encoding even at high quality levels ($n{=}20$, roughly 31.5\,dB PSNR). In comparison, DCVC-RT~\cite{jia2025towards}, the first neural video codec capable of real-time encoding, requires a high-power GPU to reach similar throughput and does not support CPU inference.

\runin{Extreme compression ratios} FRAPPE-Image can provide extreme compression ratios in excess of 5000:1, while the lowest AVIF and JPEG settings only reach 352:1 and 139:1, respectively. Among the learned baselines only WaLLoC reaches the sub-25\,dB PSNR regime FRAPPE targets at these ratios; mbt2018's quality grid bottoms at 28\,dB and is therefore absent from the lowest two PSNR regimes of Table~\ref{tab:bd_rate}.

\runin{Fixed quality target} For a fixed quality target of 21\,dB PSNR, FRAPPE-Image encodes 1.7 times faster than JPEG (915\,MP/sec vs 544\,MP/sec) while providing 22 times higher compression ratio (0.0080\,bpp vs 0.173\,bpp). AVIF's throughput on the same CPU testbed ranges from 1.97 to 6.04\,MP/sec.

\runin{PSNR/SSIM lead of mbt2018 comes at a steep throughput cost} mbt2018 retains a BD-PSNR advantage of $+2.2$ to $+4.2$\,dB over FRAPPE-Image across the $[0.1, 1)$\,bpp band (Table~\ref{tab:bd_metric}), but at 0.16--0.17\,MPx/s on the same CPU testbed---up to ${\sim}1000\times$ slower than FRAPPE's encode throughput (74--168\,MPx/s) at matched rates. The PSNR-optimal regime mbt2018 dominates is therefore unreachable in the asymmetric, on-sensor encoding setting that motivates FRAPPE.

\section{Conclusion}
We presented FRAPPE, a powerful representation-learning technique suitable for zero-overhead variable-rate lossy compression on resource-constrained sensors. Using this framework, we built a practical image compression system, FRAPPE-Image, which performs favorably against existing codecs in terms of the trade-off between rate, distortion, and encoding complexity.

\runin{Limitations and future work} The framework applies to any 1D, 2D, or 3D signal with an arbitrary channel count, but our experiments cover only RGB images; instantiations for audio, hyperspectral images, video, and 3D volumes are an obvious extension. FRAPPE-Image is intentionally biased toward low-rate, perceptual-quality operating points and the encoder-side resource budget; at moderate-to-high bitrates conventional symmetric codecs and learned baselines with heavier analysis transforms retain a rate--distortion advantage on PSNR/SSIM, and our experiments do not include hyperprior, autoregressive, or recent variable-rate learned codecs (e.g.\ conditional, prompt-tuned, or quantizer-tuning approaches)---a head-to-head against these on the same CPU testbed is left to future work. Variable-rate operation here is realized by storing one merged-decoder snapshot per supported channel count $n$ (21 snapshots in FRAPPE-Image), which is a substantial storage and deployment burden; training a single decoder with random channel dropout~\cite{hojjat2025mcucoder} to handle arbitrary channel subsets is a natural next step. Broader datasets (Tecnick, CLIC), higher resolutions, libaom-av1 with tuned speed presets, and ablations over the compander, $\rho$, and the $\lambda_m$ schedule are all left to a longer companion paper. The entropy stage (JPEG-LS over companded 8-bit latents) is deliberately simple and CPU-friendly; substituting a learned or per-image entropy model is straightforward within our four-function entropy contract and could close part of the rate gap at moderate bitrates without changing the encoder.

% \section*{Acknowledgment}
% acknowledgement

\clearpage
\bibliographystyle{IEEEtran}
\bibliography{refs}

\clearpage
\appendices
\section{Regime-Localized Bjontegaard-Delta Analysis}
\label{sec:detailed_results}
Tables~\ref{tab:bd_metric} and~\ref{tab:bd_rate} summarize the operating points of Fig.~\ref{fig:rd_throughput} from two complementary viewpoints, extended with three additional CPU-only baselines: JPEG XL (libjxl, effort=7), LiVeAction~\cite{jacobellis2026liveaction} (the published \texttt{lsdir\_f16c48} checkpoint), and MCUCoder~\cite{hojjat2025mcucoder} (the published MS-SSIM checkpoint via fp32 PyTorch; reported throughput is therefore an upper bound on the deployed INT8/CMSIS-NN encoder). Both anchor the comparison on FRAPPE-Image and prune to a single representative point per (codec, regime) pair, taken at the regime's median value of the binning axis. ``Setting'' is the codec's sweep parameter: JPEG and AVIF Pillow quality, mbt2018 and WaLLoC integer quality, and FRAPPE-Image transmitted channel count $n$. Distortion values are means across the 24 Kodak images; throughput is the median over five timed runs (one warmup) on the AMD EPYC 9354 CPU testbed of Section~\ref{sec:results}. Bjontegaard-Delta values~\cite{bjontegaard2008improvements} are computed via PCHIP interpolation on a window comprising the regime plus one immediately adjacent point on either side; entries marked ``--'' are regimes where FRAPPE and the test codec curves do not overlap sufficiently along the integration axis. FRAPPE rows are zero by construction.

Table~\ref{tab:bd_metric} bins by rate (1/3-decade bpp regimes above 0.0464\,bpp, factor $10^{1/3}\!\approx\!2.15$, with all lower-rate operating points collapsed into a single $<0.0464$ regime) and reports BD-Metric (BD-PSNR / BD-SSIM / BD-DISTS), the average distortion difference at matched rate. Positive entries indicate the test codec achieves higher quality than FRAPPE in that rate regime. Table~\ref{tab:bd_rate} instead bins by quality (PSNR in 2.5\,dB regimes from 22.5 to 32.5\,dB) and reports BD-Rate, the average percentage rate difference at matched distortion. Negative entries indicate the test codec needs less rate than FRAPPE to match the same quality. The two views are duals: each guarantees overlap along its respective integration axis by construction, eliminating the disjoint-curve regime in which BD-statistics would otherwise be undefined.

\begin{table*}[!t]
\centering
\caption{Rate-Binned BD-Metric on Kodak}
\label{tab:bd_metric}
\scriptsize
\setlength{\tabcolsep}{2pt}
\renewcommand{\arraystretch}{0.9}
\begin{tabular}{cllrrrrrrrr}
\toprule
Regime [bpp] & Codec & Setting & bpp & PSNR (dB) & SSIM & DISTS (dB) & Thr.\ (MPx/s) & BD-PSNR (dB) & BD-SSIM & BD-DISTS (dB) \\
\midrule
\multirow[t]{3}{*}{$<0.0464$} & WaLLoC & $q=2$ & 0.02035 & 21.93 & 0.5866 & 4.73 & \underline{74.02} & $-0.80$ & \underline{$-0.0475$} & \underline{$-0.51$} \\
 & LiVeAction & $q=4$ & 0.03406 & 22.98 & 0.6376 & 5.13 & 1.90 & \underline{$-0.74$} & $-0.0540$ & $-0.83$ \\
\rowcolor{black!10} & FRAPPE & $n=4$ & 0.01337 & 21.90 & 0.5870 & 4.66 & \textbf{510.48} & $\mathbf{0.00}$ & $\mathbf{0.0000}$ & $\mathbf{0.00}$ \\
\midrule
\multirow[t]{5}{*}{$[0.0464,\ 0.1)$} & AVIF & $q=1$ & 0.06814 & 25.60 & 0.7727 & 6.60 & 6.04 & $\mathbf{+0.42}$ & $\mathbf{+0.0087}$ & \underline{$-0.39$} \\
 & WaLLoC & $q=8$ & 0.06279 & 24.02 & 0.7174 & 6.12 & \underline{45.45} & $-1.05$ & $-0.0421$ & $-0.79$ \\
 & LiVeAction & $q=9$ & 0.07338 & 24.36 & 0.7309 & 6.25 & 1.72 & $-1.02$ & $-0.0466$ & $-0.91$ \\
 & MCUCoder & $q=1$ & 0.08542 & 20.55 & 0.6900 & 5.21 & 24.29 & $-4.30$ & $-0.0884$ & $-2.03$ \\
\rowcolor{black!10} & FRAPPE & $n=10$ & 0.05792 & 24.81 & 0.7520 & 6.75 & \textbf{300.73} & \underline{$0.00$} & \underline{$0.0000$} & $\mathbf{0.00}$ \\
\midrule
\multirow[t]{8}{*}{$[0.1,\ 0.215)$} & JPEG & $q=1$ & 0.17309 & 21.48 & 0.6152 & 4.56 & \textbf{543.97} & $-4.73$ & $-0.1997$ & $-3.72$ \\
 & JPEG XL & $q=5$ & 0.14307 & 25.81 & 0.7859 & 7.15 & 3.17 & $-0.39$ & $-0.0426$ & $-0.89$ \\
 & AVIF & $q=15$ & 0.12245 & 27.26 & 0.8471 & 7.78 & 4.68 & \underline{$+1.05$} & \underline{$+0.0179$} & \underline{$-0.05$} \\
 & mbt2018 & $q=1$ & 0.11021 & 28.12 & 0.8556 & 7.73 & 0.17 & $\mathbf{+2.16}$ & $\mathbf{+0.0255}$ & $-0.30$ \\
 & WaLLoC & $q=16$ & 0.11444 & 25.22 & 0.7906 & 7.20 & 30.37 & $-0.95$ & $-0.0274$ & $-0.54$ \\
 & LiVeAction & $q=16$ & 0.12398 & 25.37 & 0.7955 & 7.25 & 1.81 & $-0.96$ & $-0.0315$ & $-0.69$ \\
 & MCUCoder & $q=2$ & 0.15461 & 23.69 & 0.7784 & 6.70 & 22.69 & $-3.54$ & $-0.0793$ & $-1.85$ \\
\rowcolor{black!10} & FRAPPE & $n=13$ & 0.19661 & 27.40 & 0.8740 & 8.91 & \underline{167.76} & $0.00$ & $0.0000$ & $\mathbf{0.00}$ \\
\midrule
\multirow[t]{8}{*}{$[0.215,\ 0.464)$} & JPEG & $q=10$ & 0.32659 & 26.67 & 0.8419 & 7.74 & \textbf{529.90} & $-2.77$ & $-0.0869$ & $-2.40$ \\
 & JPEG XL & $q=20$ & 0.29262 & 28.70 & 0.8874 & 9.64 & 3.63 & $+0.02$ & $-0.0159$ & $+0.10$ \\
 & AVIF & $q=35$ & 0.30673 & 30.40 & 0.9303 & 10.88 & 4.01 & \underline{$+1.49$} & \underline{$+0.0202$} & $\mathbf{+0.89}$ \\
 & mbt2018 & $q=3$ & 0.28821 & 31.39 & 0.9301 & 10.05 & 0.17 & $\mathbf{+2.95}$ & $\mathbf{+0.0244}$ & \underline{$+0.46$} \\
 & WaLLoC & $q=36$ & 0.23783 & 26.95 & 0.8711 & 9.14 & 19.63 & $-1.16$ & $-0.0130$ & $+0.20$ \\
 & LiVeAction & $q=49$ & 0.34935 & 28.15 & 0.9072 & 10.30 & 1.73 & $-1.13$ & $-0.0115$ & $+0.08$ \\
 & MCUCoder & $q=5$ & 0.33768 & 26.12 & 0.8633 & 8.80 & 19.17 & $-3.10$ & $-0.0583$ & $-1.37$ \\
\rowcolor{black!10} & FRAPPE & $n=16$ & 0.31429 & 29.22 & 0.9108 & 10.03 & \underline{130.73} & $0.00$ & $0.0000$ & $0.00$ \\
\midrule
\multirow[t]{8}{*}{$[0.464,\ 1)$} & JPEG & $q=35$ & 0.72857 & 31.01 & 0.9478 & 12.73 & \textbf{479.66} & $-0.78$ & $-0.0080$ & $+0.36$ \\
 & JPEG XL & $q=50$ & 0.51894 & 31.31 & 0.9419 & 12.56 & 3.60 & $+1.20$ & $+0.0105$ & \underline{$+2.14$} \\
 & AVIF & $q=50$ & 0.60038 & 33.39 & 0.9658 & 13.73 & 3.33 & \underline{$+2.61$} & \underline{$+0.0257$} & $\mathbf{+2.54}$ \\
 & mbt2018 & $q=5$ & 0.63418 & 35.14 & 0.9694 & 12.88 & 0.16 & $\mathbf{+4.15}$ & $\mathbf{+0.0280}$ & $+1.62$ \\
 & WaLLoC & $q=80$ & 0.50761 & 29.60 & 0.9355 & 12.00 & 14.66 & $-0.72$ & $+0.0003$ & $+1.00$ \\
 & LiVeAction & $q=81$ & 0.56666 & 30.13 & 0.9452 & 12.46 & 1.63 & $-0.48$ & $+0.0062$ & $+1.24$ \\
 & MCUCoder & $q=10$ & 0.60788 & 27.53 & 0.9044 & 10.57 & 15.22 & $-3.28$ & $-0.0395$ & $-0.85$ \\
\rowcolor{black!10} & FRAPPE & $n=20$ & 0.72202 & 31.57 & 0.9431 & 11.57 & \underline{73.60} & $0.00$ & $0.0000$ & $0.00$ \\
\bottomrule
\end{tabular}

\par\smallskip
\begin{minipage}{\textwidth}
\footnotesize
Each codec's mean distortion difference vs FRAPPE-Image at matched bpp. Positive BD-PSNR / BD-SSIM / BD-DISTS means the test codec achieves higher quality than FRAPPE in that rate regime.
\end{minipage}
\end{table*}

\begin{table*}[!t]
\centering
\caption{PSNR-Binned BD-Rate on Kodak}
\label{tab:bd_rate}
\scriptsize
\setlength{\tabcolsep}{2pt}
\renewcommand{\arraystretch}{0.9}
\begin{tabular}{cllrrrrrrrr}
\toprule
Regime [PSNR\,dB] & Codec & Setting & bpp & PSNR (dB) & SSIM & DISTS (dB) & Thr.\ (MPx/s) & BD-Rate$_\mathrm{PSNR}$ (\%) & BD-Rate$_\mathrm{SSIM}$ (\%) & BD-Rate$_\mathrm{DISTS}$ (\%) \\
\midrule
\multirow[t]{4}{*}{$<22.5$} & JPEG & $q=1$ & 0.17309 & 21.48 & 0.6152 & 4.56 & \underline{543.97} & $+1044.7$ & $+769.8$ & $+943.1$ \\
 & WaLLoC & $q=1$ & 0.00968 & 20.43 & 0.5165 & 4.05 & 74.83 & \underline{$+54.6$} & \underline{$+59.8$} & \underline{$+44.8$} \\
 & MCUCoder & $q=1$ & 0.08542 & 20.55 & 0.6900 & 5.21 & 24.29 & $+760.1$ & -- & $+296.0$ \\
\rowcolor{black!10} & FRAPPE & $n=3$ & 0.00800 & 21.08 & 0.5455 & 4.17 & \textbf{914.94} & $\mathbf{0.0}$ & $\mathbf{0.0}$ & $\mathbf{0.0}$ \\
\midrule
\multirow[t]{5}{*}{$[22.5,\ 25)$} & JPEG & $q=5$ & 0.22115 & 23.85 & 0.7326 & 5.80 & \textbf{530.25} & $+516.0$ & $+489.1$ & $+558.1$ \\
 & WaLLoC & $q=4$ & 0.03244 & 22.79 & 0.6373 & 5.21 & 66.27 & $+55.8$ & \underline{$+51.4$} & \underline{$+61.8$} \\
 & LiVeAction & $q=4$ & 0.03406 & 22.98 & 0.6376 & 5.13 & 1.90 & \underline{$+53.9$} & $+56.9$ & $+75.7$ \\
 & MCUCoder & $q=2$ & 0.15461 & 23.69 & 0.7784 & 6.70 & 22.69 & $+361.2$ & $+129.2$ & $+225.0$ \\
\rowcolor{black!10} & FRAPPE & $n=8$ & 0.04101 & 24.05 & 0.7159 & 6.24 & \underline{415.67} & $\mathbf{0.0}$ & $\mathbf{0.0}$ & $\mathbf{0.0}$ \\
\midrule
\multirow[t]{7}{*}{$[25,\ 27.5)$} & JPEG & $q=10$ & 0.32659 & 26.67 & 0.8419 & 7.74 & \textbf{529.90} & $+165.0$ & $+168.0$ & $+196.9$ \\
 & JPEG XL & $q=5$ & 0.14307 & 25.81 & 0.7859 & 7.15 & 3.17 & $+22.6$ & $+50.1$ & $+52.4$ \\
 & AVIF & $q=5$ & 0.07497 & 25.83 & 0.7865 & 6.81 & 5.78 & $\mathbf{-27.4}$ & $\mathbf{-12.9}$ & \underline{$+16.2$} \\
 & WaLLoC & $q=25$ & 0.17037 & 26.11 & 0.8391 & 8.29 & 24.64 & $+56.8$ & $+31.7$ & $+31.1$ \\
 & LiVeAction & $q=25$ & 0.18668 & 26.31 & 0.8461 & 8.35 & 1.93 & $+57.8$ & $+35.8$ & $+39.4$ \\
 & MCUCoder & $q=6$ & 0.39355 & 26.56 & 0.8717 & 9.30 & 18.17 & $+220.2$ & $+103.3$ & $+113.2$ \\
\rowcolor{black!10} & FRAPPE & $n=12$ & 0.08027 & 25.64 & 0.7881 & 7.34 & \underline{237.35} & \underline{$0.0$} & \underline{$0.0$} & $\mathbf{0.0}$ \\
\midrule
\multirow[t]{8}{*}{$[27.5,\ 30)$} & JPEG & $q=20$ & 0.50827 & 29.14 & 0.9139 & 10.25 & \textbf{469.02} & $+61.4$ & $+60.7$ & $+47.2$ \\
 & JPEG XL & $q=20$ & 0.29262 & 28.70 & 0.8874 & 9.64 & 3.63 & $+3.3$ & $+24.7$ & $-0.1$ \\
 & AVIF & $q=25$ & 0.18705 & 28.59 & 0.8903 & 9.08 & 4.61 & \underline{$-32.6$} & \underline{$-22.2$} & $\mathbf{-19.6}$ \\
 & mbt2018 & $q=1$ & 0.11021 & 28.12 & 0.8556 & 7.73 & 0.17 & $\mathbf{-50.1}$ & $\mathbf{-23.0}$ & $+5.2$ \\
 & WaLLoC & $q=56$ & 0.35890 & 28.29 & 0.9100 & 10.71 & 20.83 & $+38.6$ & $+15.0$ & \underline{$-9.6$} \\
 & LiVeAction & $q=49$ & 0.34935 & 28.15 & 0.9072 & 10.30 & 1.73 & $+41.6$ & $+14.0$ & $-3.5$ \\
 & MCUCoder & $q=11$ & 0.66242 & 27.71 & 0.9090 & 10.81 & 14.62 & $+203.2$ & $+120.4$ & $+51.2$ \\
\rowcolor{black!10} & FRAPPE & $n=16$ & 0.31429 & 29.22 & 0.9108 & 10.03 & \underline{130.73} & $0.0$ & $0.0$ & $0.0$ \\
\midrule
\multirow[t]{7}{*}{$[30,\ 32.5)$} & JPEG & $q=40$ & 0.78557 & 31.42 & 0.9530 & 13.25 & \textbf{469.90} & $+15.6$ & $+4.3$ & $-18.9$ \\
 & JPEG XL & $q=30$ & 0.39710 & 30.04 & 0.9201 & 11.14 & 3.65 & $-22.4$ & $-17.4$ & $-34.9$ \\
 & AVIF & $q=40$ & 0.37911 & 31.25 & 0.9439 & 11.60 & 3.53 & \underline{$-43.2$} & \underline{$-42.0$} & $\mathbf{-44.2}$ \\
 & mbt2018 & $q=3$ & 0.28821 & 31.39 & 0.9301 & 10.05 & 0.17 & $\mathbf{-57.7}$ & $\mathbf{-45.4}$ & $-21.6$ \\
 & WaLLoC & $q=100$ & 0.61707 & 30.56 & 0.9501 & 13.26 & 14.19 & $+20.8$ & $-18.1$ & \underline{$-44.0$} \\
 & LiVeAction & $q=81$ & 0.56666 & 30.13 & 0.9452 & 12.46 & 1.63 & $+12.1$ & $-16.9$ & $-38.2$ \\
\rowcolor{black!10} & FRAPPE & $n=20$ & 0.72202 & 31.57 & 0.9431 & 11.57 & \underline{73.60} & $0.0$ & $0.0$ & $0.0$ \\
\bottomrule
\end{tabular}

\par\smallskip
\begin{minipage}{\textwidth}
\footnotesize
Each codec's mean percentage rate difference vs FRAPPE-Image at matched distortion. Negative BD-Rate means the test codec needs less rate than FRAPPE to reach the same quality. The PSNR column anchors the regime; SSIM/DISTS BD-Rate cells use the same PSNR-binned slice and may yield ``--'' where SSIM/DISTS do not overlap despite matched PSNR.
\end{minipage}
\end{table*}

\section{Evaluation Methodology Details}
\label{sec:eval_details}

This appendix documents harness-level choices in the open-source evaluation pipeline that materially affect the numbers in Section~\ref{sec:results} and Appendix~\ref{sec:detailed_results}.

\runin{Throughput vs rate-distortion input shape} Rate-distortion metrics use the native Kodak resolution (768$\times$512 or 512$\times$768); encoder throughput is measured on 512$\times$512 center crops, sidestepping per-codec divisibility constraints (mbt2018 requires multiples of 64) and keeping the throughput denominator constant across codecs. Input pre-staging is excluded from the timer.

\runin{Single-threaded CPU} All CPU encodes (Pillow JPEG and AVIF, mbt2018, WaLLoC, FRAPPE) run with \texttt{torch.set\_num\_threads(1)}; Pillow's libavif backend is at default speed and effort with no tile or thread tuning. Reported throughputs are per-thread.

\runin{mbt2018 bitstream} The vendored mbt2018 baseline reports bpp from forward-pass likelihoods ($-\!\sum\!\log_2 p / n_\text{pixels}$) rather than from a real bitstream---the autoregressive context-model \texttt{compress()} loop is not invoked. Likelihood-based bpp is a tight lower bound on what an entropy coder over the same likelihoods would achieve, but real CPU encode time would be substantially higher than the forward-pass throughput plotted here, since the autoregressive serialization dominates on CPU. The reported mbt2018 throughput should therefore be read as an upper bound on a deployable encoder.

\runin{WaLLoC variable-rate} WaLLoC's quality parameter is a bicubic resize-down applied inside the encoder before the wavelet and learned analysis transforms. The resize cost is included in WaLLoC's reported throughput; the bpp denominator is the original (pre-resize) pixel count, matching the user-facing rate.

\runin{FRAPPE encode timing} FRAPPE's throughput is end-to-end through encoder forward pass plus int8 quantization, device-to-host transfer of the quantized latents, and CPU-side latent arrangement plus JPEG-LS entropy coding. Each measurement is one untimed warmup epoch over the 24-image dataset followed by five timed epochs; throughput is megapixels per image divided by the median per-image total time.

\end{document}